\documentclass{article}
\usepackage{ijcai23}

\usepackage{times}
\usepackage{soul}
\usepackage{url}
\usepackage[hidelinks]{hyperref}
\usepackage[utf8]{inputenc}
\usepackage[small]{caption}
\usepackage{graphicx}
\usepackage{amsmath}
\usepackage{amsthm}
\usepackage{booktabs}
\usepackage{algorithm}
\usepackage{algorithmic}
\usepackage[switch]{lineno}
\usepackage{comment}
\usepackage{graphicx}
\usepackage{subcaption}

\usepackage{multirow}
\usepackage{multicol}
\usepackage{tabularx}

\title{Dataset Bias in Human Activity Recognition}

\author{
Nilah Ravi Nair$^1$\and
Lena Schmid$^2$\and
Fernando Moya Rueda$^4$\and
Markus Pauly$^{2,3}$\and
Gernot A. Fink$^4$\And
Christopher Reining$^1$
\and
\affiliations
$^1$Chair of Material Handling and Warehousing, TU Dortmund University\\
$^2$Department of Statistics, TU Dortmund University\\
$^3$Research Center Trustworthy Data Science and Security, UA Ruhr\\
$^4$Pattern Recognition in Embedded Systems Group, TU Dortmund University\\
\emails
\{nilah.nair, lena.schmid, fernando.moya, markus.pauly, gernot.fink, christopher.reining\}@tu-dortmund.de
}

\begin{document}

\maketitle

\begin{abstract}
    When creating multi-channel time-series datasets for Human Activity Recognition (HAR), researchers are faced with the issue of subject selection criteria.
    It is unknown what physical characteristics and/or soft-biometrics, such as age, height, and weight, need to be taken into account to train a classifier to achieve robustness towards heterogeneous populations in the training and testing data.
    This contribution statistically curates the training data to assess to what degree the physical characteristics of humans influence HAR performance. We evaluate the performance of a state-of-the-art convolutional neural network on two HAR datasets that vary in the sensors, activities, and recording for time-series HAR. The training data is intentionally biased with respect to human characteristics to determine the features that impact motion behaviour. The evaluations brought forth the impact of the subjects' characteristics on HAR. Thus, providing insights regarding the robustness of the classifier with respect to heterogeneous populations. The study is a step forward in the direction of fair and trustworthy artificial intelligence by attempting to quantify representation bias in multi-channel time series HAR data. \end{abstract}

\section{Introduction} \label{intro}

    Automatic recognition of physical activities in time series of sensor data is referred to as Human Activity Recognition (HAR).
    It is a significant research area in human-technology interaction and mobile and ubiquitous computing.
    Methods of HAR are deployed in a wide variety of application domains \cite{reining_human_2019}.
    The data varies according to the application, but it is usually multi-channel time series data from videos or on-body sensors (OBDs) such as inertial measurement units (IMUs).
    For training a robust classifier, the efficient creation of a high-quality dataset is crucial \cite{cruz-sandoval_semi-automated_2019,reining_annotation_2020,avsar_benchmarking_2021}.
    The issue of classifier robustness is multi-faceted. Recent research has shown that a classifier often inherits bias from the dataset.
    Aspects such as varying sensor placements \cite{chang_systematic_2020}, a shift of the domain \cite{khan_scaling_2018}, inconsistent labels \cite{reining_annotation_2020} and a lack of class balance \cite{niemann_lara_2020} have already been given attention to by researchers under the area of dynamic inductive biases.    
    
    Recently, representation bias in classifier or recognition models has brought forth the influence of human subjects' characteristics in deep network models. 
    However, the relevance of subject characteristics has not been explored in the case of multi-channel time series HAR datasets. 
    Each individual has a unique motion behaviour when performing various activities that can be identified with short-term signal patterns \cite{retsinas_person_2020}.
    As a result, in cases where the subjects' physical characteristics differ in the training and testing data, a decline in performance is expected.
    It is reasonable to assume that a robust classifier must incorporate training data from diverse subjects with varying physical characteristics.
    For example, when creating the LARa-dataset, the researchers took care that the subjects chosen for the laboratory recordings represent the population's demographics in the target domain \cite[p. 13]{niemann_lara_2020} -- in this case, a warehouse. 
    This approach aims to ensure that the classifier is robust towards the motion patterns of the individuals from the target domain, who are, in many cases, unknown while gathering training data. 
    Furthermore, these evaluations of the effect of subject characteristics on HAR would be the first step towards ensuring fair and trustworthy models for multi-channel time-series HAR applications.
	
    So far, it is unclear to what degree the physical characteristics of humans influence activity recognition performance.
    Researchers will be faced time and again with choosing who to record and what characteristics, such as handedness, gender or age, to take into account.
    Hence, this contribution aims to determine what physical characteristics to consider when selecting subjects for creating a robust classifier.
    In an attempt to answer this, we evaluate how far a bias of human physical characteristics in the training and testing datasets influences classifier performance -- does this influence differ between various human characteristics and activities performed? 
	
	The answers to these questions have far-reaching implications for dataset creation, particularly for crowd generation.
	In general, one would expect that people with the same physical characteristics share motion idiosyncrasies that facilitate activity recognition in homogeneous populations.
	Still, there may be physical characteristics that, despite their proven influence on human motion behaviour, have negligible impact on HAR performance.
	Possibly, it may be sufficient to integrate a random, yet sufficiently large, set of people in some cases. 
	We are dealing with these issues by deploying a series of experiments using different datasets to generalize our results as far as possible.
    
    The remainder of this contribution is structured as follows. 
    Section~\ref{rel_work} presents findings regarding the motion behaviour of humans resulting from their physical characteristics and their connection to identity and activity recognition.
    Section~\ref{method} proposes a method and explains the proposed experiments to quantify the influence of subject selection on HAR performance.
    Section~\ref{results} presents the quantitative results of the proposed experiments using different datasets.
    Section~\ref{conclusions} discusses the findings of these contributions and concludes with an outlook.

\section{Related Work} \label{rel_work}

 Recent developments in machine learning (ML) applications, such as deep face recognition, loan and credit and product suggestion applications, have highlighted the importance of evaluating datasets for potential biases \cite{van2022overcoming}. 
 Studies have shown that ML model bias can result from a bias in the underlying dataset. 
 \cite{van2022overcoming} provides an overview of the eight biases found in datasets: social, measurement, representation, label, algorithmic, evaluation, deployment and feedback bias. Recent research in FairFace has researchers looking into representation bias to obtain fair and trustworthy ML models \cite{karkkainen2021fairface}. The authors created a balanced dataset concerning age, gender and ethnic aspects. 
 It facilitates a more accurate classification model with the help of Twitter, Flickr, and the public image dataset to help ensure equal representation of each characteristic. Similar research in the area of computer vision motivated authors of \cite{wang2022revise} to create a tool called REVISE, which facilitates the detection of potential biases in a visual dataset with respect to the object, person and geography-based analysis. 
 
For HAR using multi-channel time-series data, previous research has focused on the various dynamic inductive biases.
They include the type of sensors, sensor positions, segment size and pre-processing \cite{hamidi2021human}. 
For example, \cite{zhang2022integrated} uses bias and noise correction formulas as part of IMU data pre-processing. However, what has not been extensively looked into is the bias caused by the subjects selected for HAR. Recent research in gait activity-based person re-identification suggests that each individual's motion behaviour is unique, like a fingerprint \cite{alvarez2022biometric}. Nevertheless, a variety of physical characteristics that further influence how a human stands, walks, performs activities etc., \cite{riaz2015one} discusses how a single step data can reveal the soft-biometrics of an individual, such as age, gender, and height. The authors \cite{zhang_imu_2013} explored the possibilities of identifying the robot interaction partner through the gait data obtained from a single wearable sensor attached to the pelvis.

\cite{lockhart_limitations_2014} discusses segregating the model creation into three models: personal, impersonal and hybrid. Personal models are trained and tested on the same subject's activities, while impersonal or universal models use training data from users who will not be present in the test set. Finally, the hybrid model combines the personal and impersonal models. The evaluation of these models shows that using a Random Forest show, the universal model performs the worst at 76\% accuracy. In comparison, personal and hybrid models perform better at 98\% and 95\% classification accuracy, respectively. However, personal and hybrid models may not be feasible for practical applications. Thus, evaluating the physical or soft-biometric characteristics of the individual is of interest to create a robust impersonal model of the HAR classifier. \cite{ferrari2020personalization} considers the similarity between subjects to weight the training data, such that the training data are more similar to the data of the user under test. The authors further evaluated personalization models based on the similarity between users in terms of physical attributes and/or signals patterns in \cite{ferrari2020onpersonalization}. They considered a Euclidean distance between the feature vector of two subjects based on age, weight and height and visualised a multi-dimensional scaling over physical characteristics. They experimented with their method on UNIMIB-SHAR, Mobiact and MotionSense and showed that their approach improves the classification accuracy. 

Though the literature discussed above shows the impact of subjects' individuality on HAR accuracy and introduces a method to improve classification accuracy with the help of the personalisation of the model, this does not solve the issue of mass application of HAR. Thus, it would be interesting for dataset creators to understand which aspects of the subjects' characteristics cause degradation in HAR. This work statistically evaluates training data and attempts to explain the impact of the subjects' physical characteristics.

In summary, current research suggests that each individual has a unique motion behaviour. 
Their idiosyncrasies are apparent in that they allow for recognizing their unique identities and their physical characteristics.
Yet, it is unclear how far these idiosyncrasies impact HAR performance in cases of heterogeneous populations in the training and testing dataset.

\section{Statistical Analysis of Dataset Bias in Human Activity Recognition} \label{method}

When recording human movement using on-body devices (OBD), such as IMUs, sensor biases, sensor placement, or the idiosyncrasies of the individual's motion cannot be isolated. 
For practical applications of HAR, a model than can be directly implemented is desirable. The goal is to avoid the overhead of tuning it to each individual scenario's properties such as the physical characteristics of the humans. Consequently, this work focuses on statistically evaluating the impact of the subject's physical and soft-biometric characteristics. In this section, we elaborate on the hypothesis, the network and the datasets of interest to describe the method. 
    
    \subsection{Formulation of Hypotheses}
     
     Multi-channel time series datasets such as optical Motion Capture (OMoCap) systems and OBDs record the human body's movements. Previously, researchers have focused on the biases caused by the sensor placement, type of sensor, preprocessing methods, feature extraction, and label irregularities on the HAR model. Recent research have highlighted the possibility of identity and soft biometrics recognition and re-identification of the individual from these activity recordings. However, the effect of these subject specific characteristics on the classifier model is yet to be explored. On using datasets that provide the subject's soft biometrics, age, gender and handedness and physical characteristics, such as height and weight within the recording protocol, one can evaluate the impact of these characteristics on activity recognition. We refer to the affect of physical characteristics and soft biometrics of the subjects on the activity classification performance as dataset bias on the classifier model.
     
     Here, we attempt to curate the training data based on heterogeneity measure (HM). Heterogeneity measure is a concept of statistics that discusses the non-uniformity of the qualities present within a set. As such, the curated training set used for neural network training will consist of different levels of heterogeneity based on the physical and soft biometric characteristics of the subjects. We hypothesis that there will be a performance variation of the classifier trained with data from different heterogeneity levels. Based on intuition, we hypothesis that as the heterogeneity of the physical characteristics of the subjects increase within a given training set, the performance measure of the activity classifier model would improve on test set of unknown subjects with varied physical characteristics.  
     
    \subsection{Classifier and Datasets}
    
        This work uses the state-of-the-art neural network for HAR, CNN-IMU network proposed by \cite{Moya2018-CNN} for its evaluation. Unlike classical classifiers, which require hand-crafted features, the convolutional layers perform the necessary feature extraction on the input data before the classification process during supervised learning. As a result, the method is robust against feature extraction biases. Furthermore, compared to the latest HAR models, such as transformers, the method is simple and does not require extensive data for training \cite{kim2022inertial}. 
        
        Here, we are interested in analysing the affect of the physical and soft biometric features prioritised by the CNN model as an effect of representation bias in the training data. The chosen network, CNN-IMU, performs efficiently, with a classification accuracy greater than 80\% on HAR datasets \cite{Moya2018-CNN}. The network is interesting due to the late fusion method, which facilitates local feature extraction per limb.
        
        For generalising the findings, the hypothesis is tested on two publicly available datasets, namely the LARa \cite{niemann_lara_2020} and MotionSense \cite{Malekzadeh:2019:MSD:3302505.3310068} datasets. The datasets were chosen based on the availability of the physical characteristics of the subjects, presence of more than ten subjects with diverse characteristics, and the public availability of the dataset to promote reproducible results.
        The two datasets differ in the type of activity, sensors used, sampling frequency and other data characteristics. LARa consists of activity recordings from 14 subjects with five IMUs placed on the subject, recorded at 100 Hz and OMoCap data recorded at 200 Hz. The activities are based on logistics order-picking scenarios. The MotionSense dataset includes 24 participants, where the subjects wore a single IMU in their front pocket. The subjects perform activities of daily living such as walking, jogging and sitting. These are recorded at 50 Hz. In both datasets, the individual characteristics of the subjects, such as age, height, weight and gender, are given. This facilitates the study of the impact of the individual's characteristics on HAR. 

    \subsection{Experimental Design}
        
Both LARa and MotionSense datasets provide the age, gender, height and weight of each subject. During the initial analysis, we found a significant correlation between the height and weight characteristics to the gender of the individuals in both datasets. Table~\ref{tab:freqHeightWeight} shows the frequency of the different characteristic values, namely, height and weight, to gender. Height and weight are classified as Short/Tall and Light/Heavy based on the datasets' median. The table shows the division after combining both datasets.

\begin{table}[h]
    \centering
       \caption{Frequencies of gender and categorised weight and height for both datasets.}
    \label{tab:freqHeightWeight}
    \begin{tabular}{c|cc}
    \textbf{Gender} & \multicolumn{2}{c}{\textbf{Weight}}\\
\hline
& Light&Heavy\\\hline \hline
Female & 12 &5\\
Male & 7 &14\\
\end{tabular}
\quad
\begin{tabular}{c|cc}
\textbf{Gender} & \multicolumn{2}{c}{\textbf{Height}}\\
\hline
& Short&Tall \\\hline \hline
Female & 16 & 1\\
Male &3  & 18\\
\end{tabular}
\end{table}

This implies that the inclusion of these characteristics in the selection of the subjects would essentially be a repetition of the trend. As a result, we focus on the age and gender of the subjects to test the hypothesis. Handedness, as a characteristic, was not included in these evaluations as information about the same was not available for the MotionSense dataset.
On further analysis of the selected characteristics, namely age and gender, age characteristic was further divided into the sub-classes young and old using the dataset-specific medians. Both datasets follow binary sub-classes of gender as male or female. 
Table~\ref{tab:freqGender} shows the frequency of the two characteristics in the LARa (left) and MotionSense (right) datasets.

\begin{table}[h]
    \centering
       \caption{Frequencies of gender and categorised age for LARa (left) and MotionSense (right) datasets.}
    \label{tab:freqGender}
    \begin{tabular}{c|cc}
    \multicolumn{3}{c}{\textbf{LARa}}\\
    \textbf{Gender} & \multicolumn{2}{c}{\textbf{Age}}\\
\hline
& Young & Old\\\hline \hline
Female & 4 & 3\\
Male & 3 & 4\\
\end{tabular}
\quad
\begin{tabular}{c|cc}
   \multicolumn{3}{c}{\textbf{MotionSense}}\\
\textbf{Gender} & \multicolumn{2}{c}{\textbf{Age}}\\
\hline
& Young & Old\\\hline \hline
Female & 7 & 3\\
Male & 6  & 8\\
\end{tabular}
\end{table}

Thus, we have four different combinations of these two characteristics: young woman, old woman, young man, and old man.
As seen from the table, the LARa dataset has an almost similar number of subjects in age and gender sub-classes. However, more significant variations in the number of subjects can be found in MotionSense characteristics sub-classes. 

Four subjects are used for training the activity classifier for each dataset, and the rest are used for testing. The heterogeneity in the training sample is determined by the number assigned based on different characteristic values, as shown in Table ~\ref{tab:HM}. For example, '1' implies that the selected training set subjects have no heterogeneity. As a result, all subjects in the sample have the same characteristic values. 
Similarly, '4' implies that the training set is heterogeneous, and all subjects have different characteristics. 

\begin{table}[h!]  
\caption{Description of the Heterogeneity Measure for the Training Data.}  
\label{tab:HM}
    \begin{tabular}{c|l}
        \textbf{Group} & \textbf{Heterogeneity Measure}  \\
        \hline
        \hline
        1 &  all subjects have the same characteristics\\
        2 & \multicolumn{1}{p{7cm}}{two subjects with similar characteristics (e.g. two old women and two young men are used in the training sample)} \\
        3 & three subjects with varied characteristics\\
        4 & all subjects have different characteristics
    \end{tabular}
 \end{table}   
 
The HM '2' can be further divided into two subgroups depending on how the two types of subjects differ: (2a) refers to differences in one characteristic (e.g. two young men and two young women are used in the training sample), and (2b) refers to the difference in both characteristics (e.g. one young woman and three old men in the training sample). 

To avoid selection bias, the subjects for the training sample were randomly selected based on the HM of the experiment. As a result, ten different experiments were obtained for each heterogeneity group. However, for the two extreme cases (HM '1' and '4'), it is not possible to have ten different training samples. As a result, experiments were conducted on the maximum number of training sample sets feasible on the respective dataset. This resulted in 27 different settings for the LARa and 28 for MotionSense. A detailed description of the settings can be found in Tables~\ref{tab:ExpLARa} and~\ref{tab:ExpMS} in the appendix.

\subsubsection{Training Procedure}

The weights of the network are initialised using the orthogonal initialisation method. The Cross-Entropy Loss function is utilised to calculate activity classification loss. The Root Mean Square Propagation (RMSProp) optimisation is used with a momentum of $0.9$ and weight decay of $5\times10^{-4}$. Gaussian noise with mean $\mu = 0$ and standard deviation (SD) $\sigma = 0.01$ is added to the sensor measurements to simulate sensor inaccuracies \cite{moya2018convolutional}\cite{grzeszick2017deep}. Dropout of probability $p=0.5$ was applied on the Multi-layer Perceptron (MLP), and early-stopping was implemented to avoid over-fitting.

\subsubsection{Evaluation Metric}

To measure the activity metrics, accuracy and weighted F1 score (wF1) were considered. wF1 was considered for evaluation due to the unbalanced nature of the activity recordings in the datasets. In addition, the standard deviation of the average accuracy and average wF1 was considered in cases where multiple trials of the training-test were considered. Analysing the standard deviation (SD) for these performance measurements would be support in uncertainty quantification. 
\section{Experiments and Results} \label{results}

As discussed in the previous section, to analyse the impact of particular human features on activity recognition, the subjects of the training and validation sets were chosen based on statistical hypothesis. 
The appropriate hyper-parameters of the neural network were determined based on the evaluations of previous works and empirical studies.

In the case of the LARa dataset, we considered the oMoCap data for experimentation to ensure the inclusion of all 14 subjects. The oMoCap recordings were downsampled from 200 Hz to 100 Hz to reciprocate the frequency of the LARa IMU dataset. A window size of 100 frames with a step length of 12 was considered the best window size as per the evaluations of \cite{Moya2018-CNN} and \cite{niemann_lara_2020}. Further, a batch size of 100 and a learning rate of 0.0001 with epoch 32 were found to provide 77.63\% accuracy and wF1 of 78.77\%. For the MotionSense dataset, we obtained 95.6\% accuracy and wF1 of 95.54\% when trained and tested on all subjects for activity recognition on a CNN-IMU network with a single block of four convolutional layers. A single block of convolutional layers was considered as the dataset consisted of a single sensor placed on the chest with nine channels. All six activities were evaluated for classification. A window size of 200 samples with a step size of 25 was found to achieve high accuracies in the empirical studies. A learning rate of 0.001, epoch 20 and batch size 65 was found to provide reasonable accuracy measures on the overall train-test set. To evaluate consistency in the experimentation, we use a fixed seed for initialising the neural network. In addition, an experiment on training data was performed five times to evaluate the consistency of the result. Standard deviation analysis on these values would provide a perspective on the uncertainty measure of these evaluations. 

\begin{figure}[h]
\centering
   \includegraphics[width=1\linewidth]{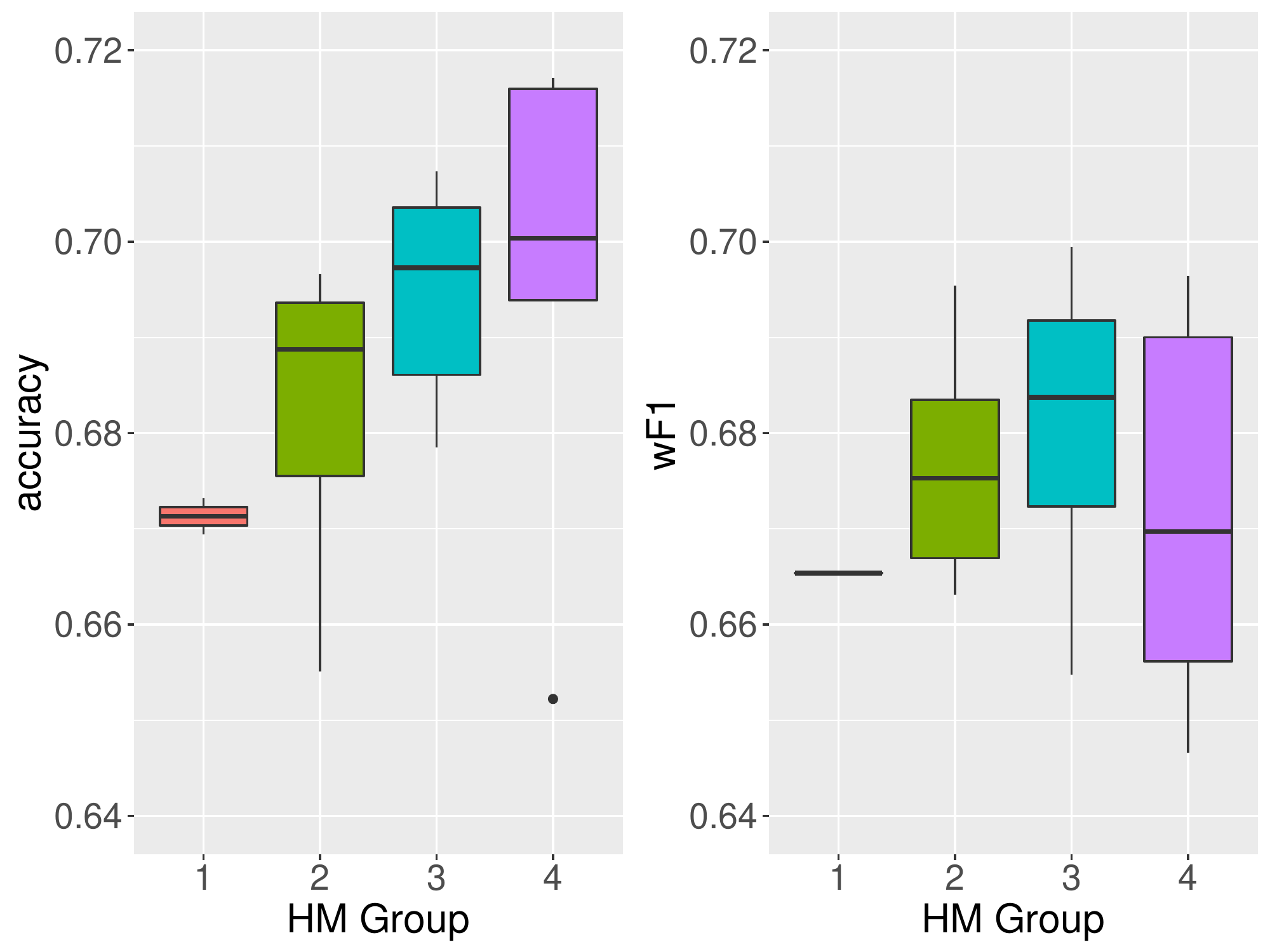}
   \caption{Results on the average accuracy (left) and average wF1 (right), measured in percentage, for the LARa dataset on all HM groups.}
   \label{fig:Lara}
\end{figure}

\begin{figure}[h]
\centering
   \includegraphics[width=1\linewidth]{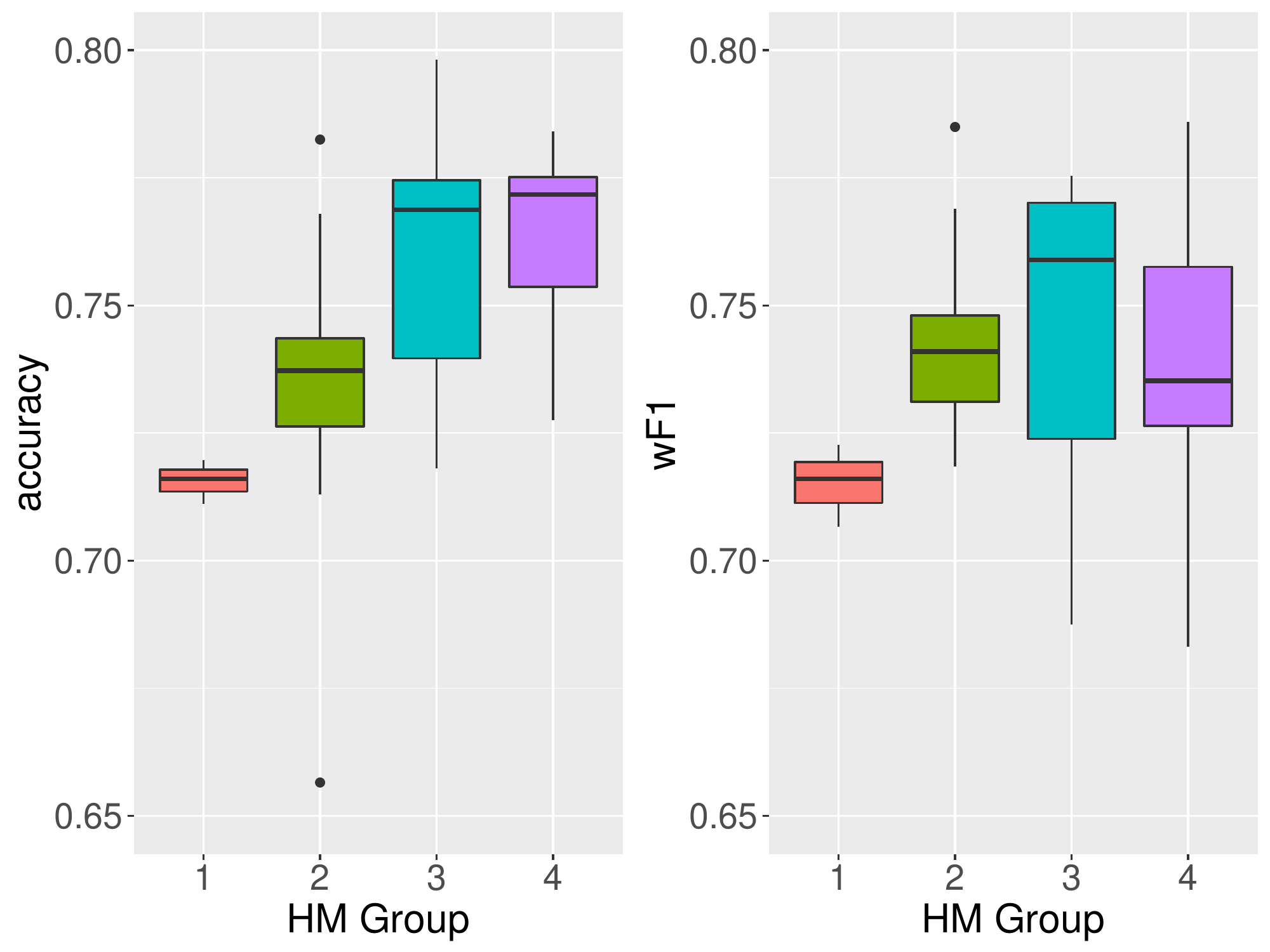}
   \caption{Results on average accuracy (left) and average wF1 (right), measured in percentage, for the MotionSense dataset on all HM groups.}
   \label{fig:MS}
\end{figure}

Figure~\ref{fig:Lara} and Figure~\ref{fig:MS} present the boxplots of accuracy and wF1-score for each HM group for LARa and MotionSense datasets. The HM groups are given a designated colour and is followed for all plots in this work. 


The performance measures of both datasets present similar trends. The average accuracy of the classification experiments show an increase with an increase in heterogeneity measure, in alignment to our hypothesis. A similar trend is visible with the wF1 values of HM '1', '2' and '3'. However, a drop in wF1 is found in the datasets for the HM group '4'. wF1 accounts for the unbalanced activity data. As a result, we believe that the drop in wF1 in the heterogeneity group '4' may be attributed to the varied physical characteristics, influencing the network's ability to learn the movement of different activities. On closer evaluation of the number of training sequences to the wF1 for HM '4', the wF1 was found to increase proportional to the number of training sequences for the LARa dataset. However, MotionSense showed the opposite trend. The network noticeably misclassified three activity classes: walking upstairs, downstairs, and walking for the MotionSense dataset. In the case of LARa, handling center activity has a higher number of windows in the dataset. We see a general tendency of the network to classify all activities to handling center, except for synchronisation, which was predominately classified under handling upwards. This essentially shows that when the variations of the subject's physical characteristics increase, the network begins to focus on the motion of the activities and overfits on the activity class with the most number of training windows. This brings forth the well-researched issue of intra- and inter-class variability of human actions \cite{Moya2018-CNN}.

\begin{figure}[h]
\centering
   \includegraphics[width=1\linewidth]{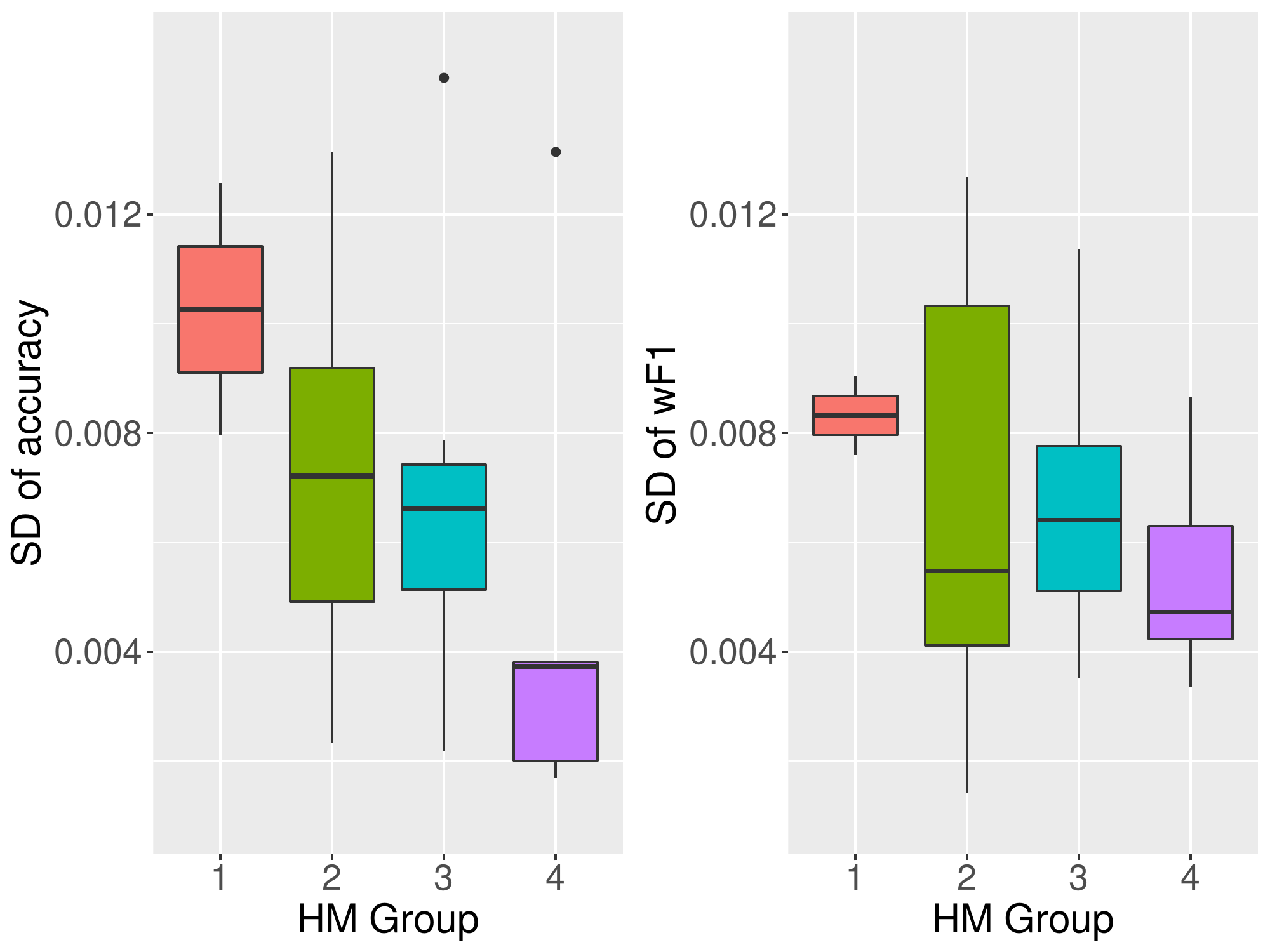}
   \caption{SD of the average accuracy (left) and average wF1 (right) for the LARa dataset on all HM Groups.}
   \label{fig:LaraSD}
\end{figure}

\begin{figure}[h]
\centering
   \includegraphics[width=1\linewidth]{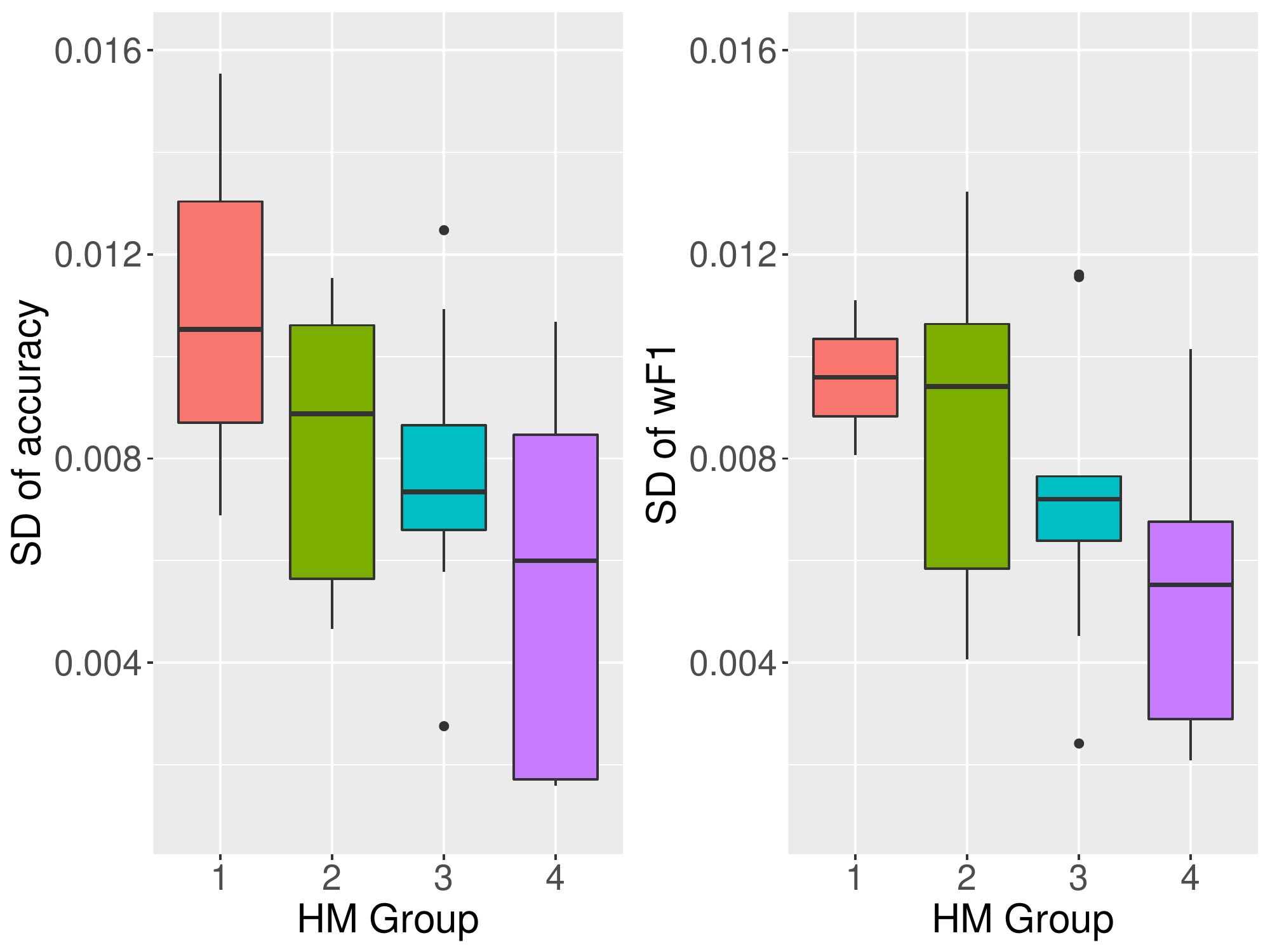}
   \caption{SD of the average accuracy (left) and average wF1 (right)) for MotionSense dataset on all HM Groups.}
   \label{fig:MSSD}
\end{figure}

The boxplots of the SD of accuracy and wF1 can be seen in Figure~\ref{fig:LaraSD} and Figure~\ref{fig:MSSD}. 


\begin{figure}[h]
\centering
   \includegraphics[width=1\linewidth]{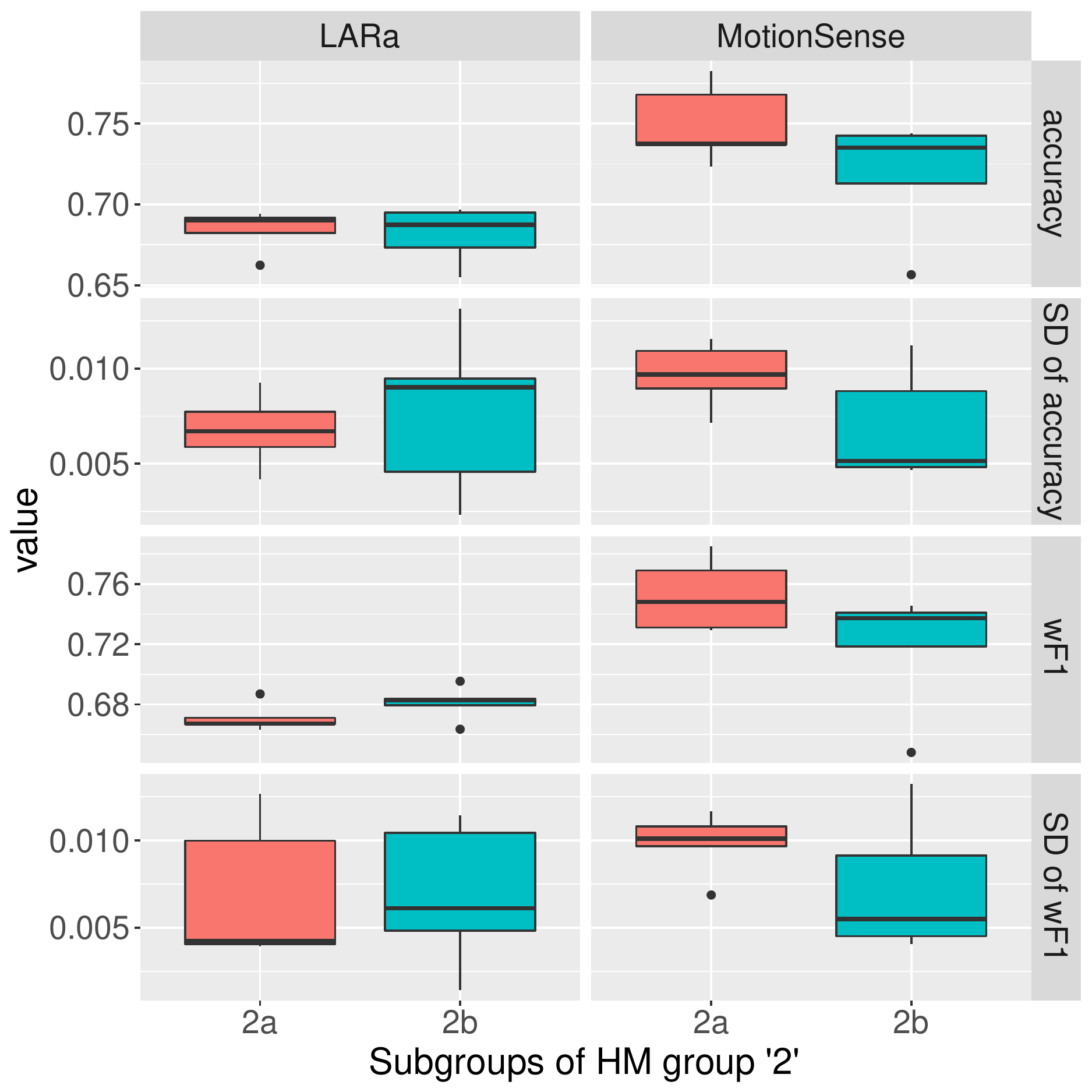}
   \caption{Evaluation of HM subgroups '2a' and '2b' for LARa and MotionSense datasets.}
   \label{fig:subgroups}
\end{figure}

Both datasets show a similar trend in the SD of average accuracy and average wF1. With increasing heterogeneity in the training sample, the SD decreases for both accuracy and wF1. This implies that, as heterogeneity measure increases the uncertainty decrease in the performance metrics. However, an anomalous trend can be seen for the SD of average wF1 for both LARa and MotionSense datasets for the HM group '2'. Here, an increase in the variation of SD can be noticed.
This difference can be attributed to the different subgroups '2a' and '2b' of HM '2'. The subgroups are based on how many classes of characteristics vary between the two sets of selected characteristics. A detailed presentation of the results based on the subgroups is presented in Figure~\ref{fig:subgroups}.

As discussed in Section~\ref{method}, '2a' refers to heterogeneity grouping where one of the characteristics, for example, gender, remains the same for all the train subjects, while the other characteristic, in this example - age, would be varied. 
This grouping brings forth the impact of the unvarying characteristics on the test set. 
However, one or two characteristics may vary in the case of '2b'. 
At least two of the subjects in the training set are retained to have similar characteristics. 
This means that the training data of HM group '2b' has more diversity in the scope of the subject's physical characteristics than '2a'. Based on the trend seen in Figure~\ref{fig:Lara} and Figure~\ref{fig:MS}, an improvement in the performance measure would be expected with increasing heterogeneity in the training sample. 
However, the LARa and MotionSense datasets show opposite trends for the heterogeneity sub-groups. 
In the case of LARa, the increase in variations seem to be favourable, thus increasing the average accuracy and wF1 values for HM group '2b'. 
However, the SD of the average accuracy and wF1 shows an increasing trend concerning the HM '2a' group. With respect to MotionSense, HM group '2a' shows a higher average accuracy and wF1 in comparison to HM group '2b'. Similar to LARa, MotionSense also shows high variations in the standard deviation of average accuracy and wF1 values. It is impossible to come to a conclusion on the trend presented by the heterogeneity subgroups '2a' and '2b'. This observation might be an effect of physical characteristics that are unaccounted for by the dataset creators and thus, unaccounted for within these experimentation. Thus, motivating the importance of evaluating the biases learned by the model from the dataset. 
\section{Conclusions} \label{conclusions}

In this work, we statistically curated the training data provided to a state-of-the-art HAR classifier to evaluate the impact of recorded subjects' physical characteristics on the network's activity classification. 
The evaluations were performed on two datasets of varied characteristics. 
Training data consisting of diverse physical characteristics was found to provide better accuracy on unseen testing data with subjects of varied physical characteristics. 
Furthermore, systematically increasing the training sample size with subjects with varied characteristics was found to improve accuracy further and decrease the standard deviations of the trials. 
Thus, we recommend dataset creators to ensure the presence of subjects with extreme characteristics in their dataset. 
The number of subjects and the diversity of their physical characteristics can thus be incremented to improve classifier performance when the effort for recording more data is reasonable. 

The datasets used in this work showed correlation between height and weight characteristics to gender. 
This might be the bias of the selected datasets. 
As a result, datasets with a larger number of subjects of varied characteristics needs to be created and then analysed as part of future work.
An anomalous trend was identified within the subgroups of heterogeneity group '2'. While the other heterogeneity groups showcased similar trends for both datasets, opposite trends were found in the case of heterogeneity '2' sub-groups of the two dataset. Acknowledging that this anomaly may be an effect of a dataset bias pertaining to physical characteristics that was unaccounted in these experiments, we urge the community to further explore dataset bias.

This work focused on binary classes within characteristics. 
As future work, evaluations on multiple sub-classes within each physical characteristic and the evaluation's impact must be performed to further generalize the conclusions of this contribution.
The influence of other classifiers needs be investigated, as well. 
The impact of a dataset bias on recognition performance may differ for new models of HAR classifiers, such as transformers and conformers. These require extensive amount of data for learning HAR through supervised learning. Thus, larger, well-documented datasets with variations in subjects' physical characteristics are required for analysis on these models.
In addition, datasets consisting of detailed subject's characteristics is desirable for identifying new dimensions of the dataset bias. For example, the impact of handedness.

\appendix

\section*{Ethical Statement}

There are no ethical issues.

\section*{Acknowledgment}

The work in this publication was supported the German Federal Ministry of Education and Research (BMBF) in the context of the project ``LAMARR Institute for Machine Learning and Artificial Intelligence'' (Funding Code: LAMARR22B), and Deutsche Forschungsgemeinschaft (DFG) in the context of the project Fi799/10-2 ``Transfer Learning for Human Activity Recognition in Logistics''.

\section*{Appendix}

The following two tables list the different training sets based on the heterogeneity measure for the two datasets, LARa and MotionSense. From the Table~\ref{tab:ExpLARa} and Table~\ref{tab:ExpMS}, it can be seen that the number of training sequences for LARa is greater than that of MotionSense. However, we see that the training sequences are proportional to the number of channels/sensors present for the respective dataset.

\begin{table}[h!]   
\caption{Detailed list of the 27 different experimental settings for the LARa dataset.}
    \label{tab:ExpLARa}
    \centering
    \begin{tabular}{ccc}
        \textbf{HM} & \textbf{Training Subsets} & \textbf{No: of Training Seq.}\\ \hline \hline 
    \multirow{2}{*}{1} & 3, 9, 10, 14& 86878\\
                       & 2, 6, 12, 13& 81832\\ \hline
    \multirow{5}{*}{2a} & 4, 6, 8, 11& 76085 \\
                        & 6, 8, 11, 12& 68069\\
                        & 2, 6, 11, 13& 78297\\
                        & 1, 5, 7, 13& 88367\\
                        & 2, 5, 6, 12& 79907\\ \hline
    \multirow{5}{*}{2b} & 2, 6, 9, 13& 90514\\
                        & 6, 9, 12, 14& 82352\\
                        & 1, 5, 6, 7& 79514\\
                        & 2, 7, 12, 13& 88523\\\hline
     \multirow{10}{*}{3}& 5, 7, 13, 14& 89371\\
                        &2, 8, 11, 14& 77781\\
                        &5, 9, 10, 11& 72610\\
                        &2, 8, 9, 11& 76757\\
                        &2, 7, 9, 14& 91287\\
                        &5, 7, 8, 9& 87134\\
                        &1, 4, 10, 14& 86959\\
                         &4, 5, 7, 14&88534\\
                       &2, 7, 10, 12& 78479\\
                        &3, 5, 9, 12& 78520\\\hline
        \multirow{5}{*}{4}& 6, 7, 10, 11 & 74938\\
                          &5, 6, 8, 10& 84112\\
                          & 1, 6, 8, 10& 85886\\
                          &3, 5, 11, 13& 75156\\
                          &2, 3, 7, 11&77319
    \end{tabular}
\end{table}
\begin{table}[h!]   
\caption{Detailed list of the 28 different experimental settings for the MotionSense dataset.}
    \label{tab:ExpMS}
    \centering
    \begin{tabular}{ccc}
    \textbf{HM} & \textbf{Training Subsets} & \textbf{No: of Training Seq.}\\ \hline \hline 
    \multirow{3}{*}{1} & 4, 9, 12, 14 & 17116 \\
                           & 3, 5, 8, 23 & 17665 \\
                           & 6, 11, 20, 21& 18478\\ \hline
    \multirow{5}{*}{2a} & 3, 17, 23, 24& 17168\\
                        & 1, 7, 10, 16& 19099 \\
                        & 7, 10, 16, 22& 18616\\
                        & 7, 9, 12, 22& 17481\\
                        & 2, 9, 12, 21& 18554\\ \hline
    \multirow{5}{*}{2b} & 3, 5, 14, 24& 17066 \\
                        & 5, 12, 14, 23& 16628\\
                        & 6, 11, 16, 20& 18208\\
                        & 6, 7, 17, 21& 18660\\
                        & 4, 15, 22, 23& 17355\\\hline
    \multirow{10}{*}{3} & 5, 10, 12, 13& 16620 \\
                        & 10, 16, 17, 19& 19398\\
                        & 6, 10, 14, 15& 17890 \\
                        & 7, 17, 14, 21& 18595 \\
                        & 8, 16, 15, 18& 19068 \\
                        & 10, 15, 16, 19 & 19699 \\
                        & 8, 15, 16, 18& 19068 \\
                        & 6, 8, 10, 16& 18674 \\
                        & 2, 5, 7, 21& 18794\\
                        &7, 10, 11, 22& 18146\\ \hline 
    \multirow{5}{*}{4} & 8, 10, 11, 15&18367\\
                        & 4, 6, 8, 10& 17956\\
                        & 10, 13, 17, 23& 16967\\
                        & 1, 6, 16, 19& 16967\\
                        & 7, 9, 11, 18& 18463\\

    \end{tabular}
\end{table}

\newpage
\bibliographystyle{named}
\bibliography{bibliography}

\end{document}